\newcommand{\be}{\begin{equation}}
\newcommand{\ee}{\end{equation}}

\documentclass[]{aa}
\usepackage{epsfig}  
\begin{document}
%
   \title{NGC~1399: a complex dynamical case}
%
%
%
   \author{N.R.Napolitano \inst{1,2}
   \and M.Arnaboldi \inst{1}
   \and M.Capaccioli \inst{1,2}}
%
%
   \institute{Osservatorio Astronomico di Capodimonte, via Moiariello 16,
              I-80131 Napoli, Italy
\and
              Dipartimento di Scienze Fisiche, Universit\`a\ di Napoli 
              Federico II,
              complesso Monte S.Angelo , via Cintia, I-80126 Napoli, Italy
		}
   \date{Received: ; accepted:}
   \titlerunning{Stellar dynamics in NGC~1399: a new analysis}
   \authorrunning{Napolitano et al.}

\abstract{
Evidence for a disturbed velocity structure in the outer regions of the galaxy NGC~1399 comes from a re-analysis of the planetary nebulae data from Arnaboldi et al. (1994). We find a strong rotation along a P.A.=140$^{\circ}$ for $R\le140''$ which is followed by a rapid drop off at larger radii, where the velocity dispersion starts to increase. This kinematical behavior can be interpreted as an indication for an interaction scenario. Interaction is advocated in previous analysis of the halo regions of this system, based on different dynamical tracers such as globular clusters and X-rays, but in all these studies the mass distribution is derived under the equilibrium hypothesis, which is not appropriate when an interaction takes place. Here we attempt a non-equilibrium dynamical analysis of NGC~1399: with a simple model and under the impulse approximation, we show that the observed kinematics is consistent with an energy injection caused by a flyby encounter of NGC~1399 with the nearby system NGC~1404. In this approach, we find a mass-to-light ratio, M/L$_\mathrm{B}$=26 $M_\odot/L_\odot$ within $R=400''$, which is about half of that requested when equilibrium is assumed, i.e. M/L$_\mathrm{B}$=56 $M_\odot/L_\odot$.
}
\maketitle
\section{Introduction}
Inner parts of galaxy clusters are the birthplaces of complex dynamical situations involving interactions between galaxies. Morphological segregation (Dressler \cite{dres80}, Whitmore et al. \cite{whit93}, Dressler et al. \cite{dres97}), HI deficiency of cluster spirals (Haynes et al. \cite{hay84}; Solanes et al. \cite{sola01}), extended envelopes in density profiles of cD galaxies (Schombert \cite{scho86}), structures in the density and temperature maps of X-rays (Roettiger et al. \cite{roet95}; Irwin \& Sarazin \cite{irw96}; Kikuchi et al. \cite{kiku00}), are some of the signatures of environmental effects on cluster galaxies. These interactions may also produce disturbed velocity structures in the outskirts of cluster galaxies, which is a well established fact for spirals. Significant statistics based on extended rotation curves measured by coupling HI with HII emission have disclosed a decline at large galactocentric distances in contrast with the asymptotic flat behavior for field spirals (Whitmore et al. \cite{whit88}; Adami et al. \cite{ada99}; Rubin et al. \cite{rub99}) and/or asymmetries (Dale et al. \cite{dale01}). On the contrary, the situation for ellipticals is still unclear owing to the lack of classical kinematical tracers in the outer regions of these galaxies.\\ 
A recent development has been offered by the first measurements of the discrete radial velocity fields of globular clusters (GCs) and planetary nebulae (PNe) in nearby early-type galaxies.
In particular, PNe follow the stellar light distribution (Ciardullo et al. \cite{ciar89}; Ciardullo et al. \cite{ciar91}; McMillan et al. \cite{mcmil93}; Ford et al. \cite{ford96}) and share the same kinematical behavior where they overlap with the integrated stellar light data (Arnaboldi et al. \cite{arn94}, \cite{arn96} and \cite{arn98}; Hui et al. \cite{hui95}). Thus, PNe appear as the natural candidates to gauge the stellar kinematics in the outskirts of ellipticals.    
Up to now the limiting factor in the use of discrete radial velocity fields has been the small number statistics (SNS). In distant galaxies (D$>$15 Mpc) 
the measured samples typically consist of less than a hundred radial velocities both for PNe and GCs, and some doubts may arise on the mass and 
angular momentum estimates based on these data alone.
Recently Napolitano et al. (\cite{napo01}, NAFC hereafter) have shown that, even under SNS regime, discrete tracers do carry some important kinematical information and can be successfully used for mass and angular momentum estimates and search for kinematical signatures of interactions. 

An interesting target in this context is NGC~1399, the cD galaxy of the Fornax cluster. Despite its apparent regular morphology, this object exhibits some peculiarities which indicate that its real dynamical status is still unclear.  
The extended stellar envelop (Schombert \cite{scho86}) and the unusual overabundance of globular clusters (Kissler-Patig et al. \cite{kisl99} and references therein), together with an increasing velocity dispersion in the outer regions from discrete radial velocity field of both GCs (Grillmair et al. \cite{grill94}, Minniti et al. \cite{minn98}, Kissler-Patig et al. \cite{kisl99}) and PNe (Arnaboldi et al. \cite{arn94}), suggest an interaction scenario, which is also supported by X-ray data. The X-ray temperature profile of the extended gaseous halo (Ikebe et al. \cite{ike96}; Rangajan et al. \cite{rang95}; Jones et al. \cite{jon97}) is compatible with the velocity dispersion from PNe and GCs. Recently, Paolillo et al. (\cite{paol01}) analysed deep ROSAT HRI data with an adaptive smoothing technique. They found: 1) an extended and asymmetric hot gas halo distributed out to the cluster scale; 2) a multicomponent density profile which flattens at 1 arcmin with one ``shoulder'' at $R=450''$, i.e. within the separation between NGC~1399 and NGC~1404; 3) filamentary structures and holes in the intensity map; 4) a displacement of the gas centroid with respect the luminous component. All these features are qualitatively expected in a tidal interaction scenario (D'Ercole et al. \cite{derc00}; Barnes \cite{bar00}).
In this context the dynamical description of this system is complex : interactions are associated to non-equilibrium situations which are difficult to handle.
Previous studies assumed equilibrium to infer the mass distribution of NGC~1399 (Bicknell et al. \cite{bick89}; Arnaboldi et al. \cite{arn94}; 
Minniti et al. \cite{minn98}; Kissler-Patig et al. \cite{kisl99}; Saglia et al.
\cite{sagl00}). 
They all show an increasing mass-to-light ratio ($M/L\geq 40 M_{\odot}/L_{\odot}$) outwards. These values match the local
M/L ratios inferred for clusters from X-ray measurements, and are consistent with those derived from the equilibrium analysis of the temperature profile of the X-ray gas around NGC~1399 
(Killeen \& Bicknell \cite{kill88}; Rangajan et al. \cite{rang95}; Jones et al. \cite{jon97}).
The question is whether these mass estimates are realistic: non-equilibrium kinematics can mimic an overall higher mass content and cause mass overestimates (Mihos \cite{miho00}).
Here we attempt a quantitative non-equilibrium analysis to derive the mass distribution of NGC~1399.\\
This paper is organised as follow. New estimates of the PNe kinematical quantities in the outer halo of NGC~1399 are obtained in Sect. 2, where some evidence of a disturbed rotational structure is also given. In Sect. 3 the PNe rotation curve and velocity dispersion profile are combined with the stellar and GCs kinematics to derive the mass distribution via the inversion of the Jeans equations, under equilibrium hypothesis. We then investigate whether NGC~1399 is a system out of equilibrium because of ongoing encounters with nearby systems in the core of the Fornax Cluster and derive a new mass accordingly. Conclusions are presented in Sect. 4. Throughout this paper, we will assume a distance of 17 Mpc for NGC~1399 (McMillan et al. \cite{mcmil93}).

\section{Kinematical analysis}
\begin{figure}
\centering
\epsfig{figure=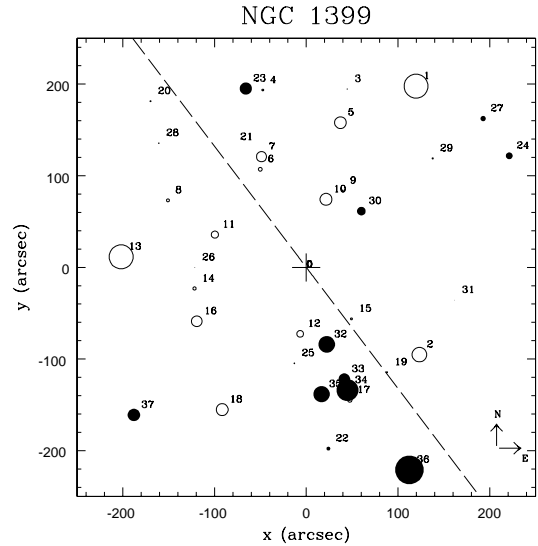,width=8cm,height=8cm}
\caption{\footnotesize PNe radial velocity field of NGC1399. Full dots are velocities above the mean velocity, $\overline{v}$, open dots are below $\overline{v}$. The dashed line is the direction of maximum gradient, i.e. the kinematical major-axis, derived from the whole sample of 37 PNe (see discussion in the text). Identification numbers are drawn; from Arnaboldi et al. (1994).}
\label{field}
\end{figure}

\begin{figure*}
\centering
\epsfig{figure=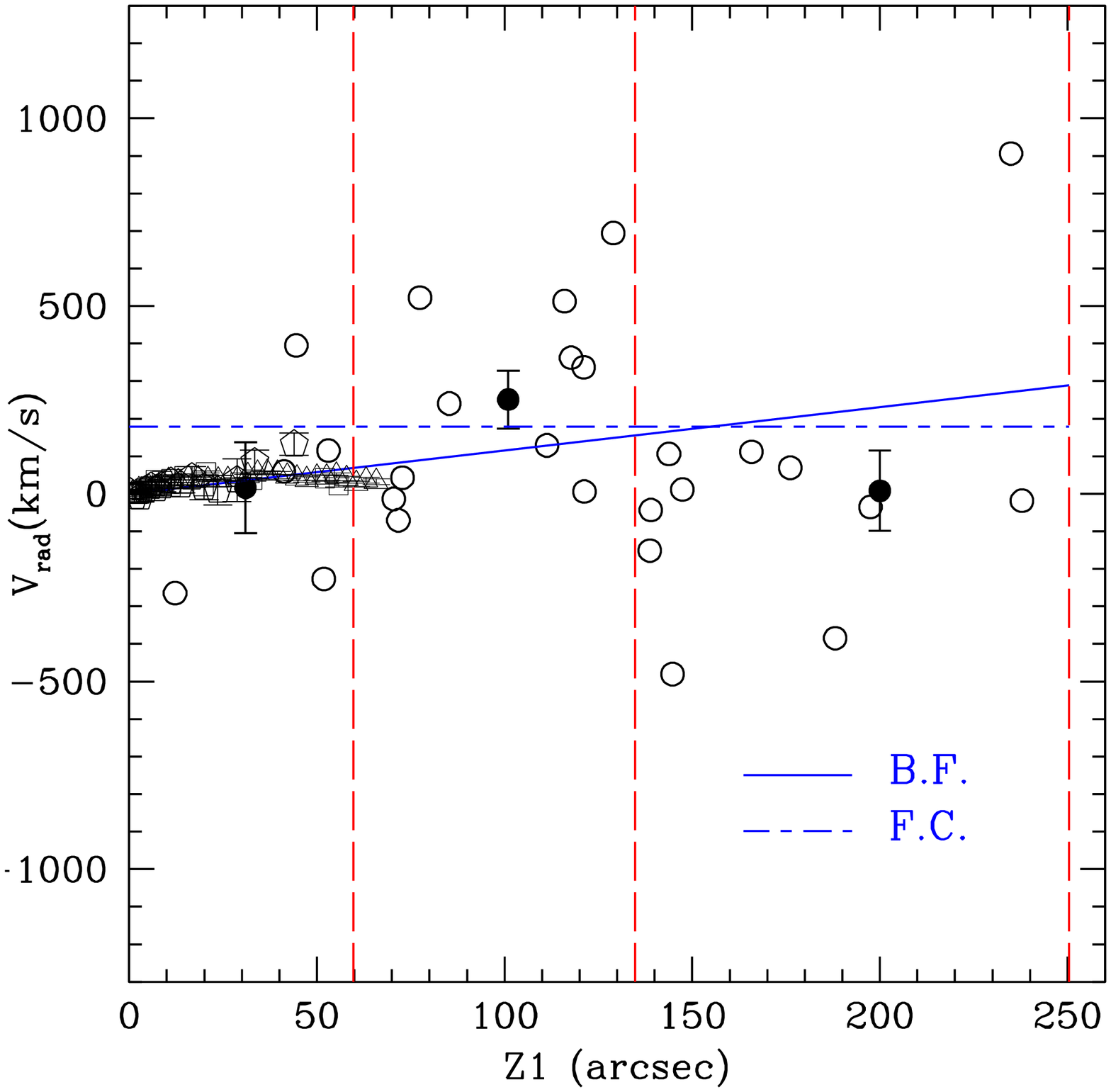,width=6.5cm,height=6.5cm}
\epsfig{figure=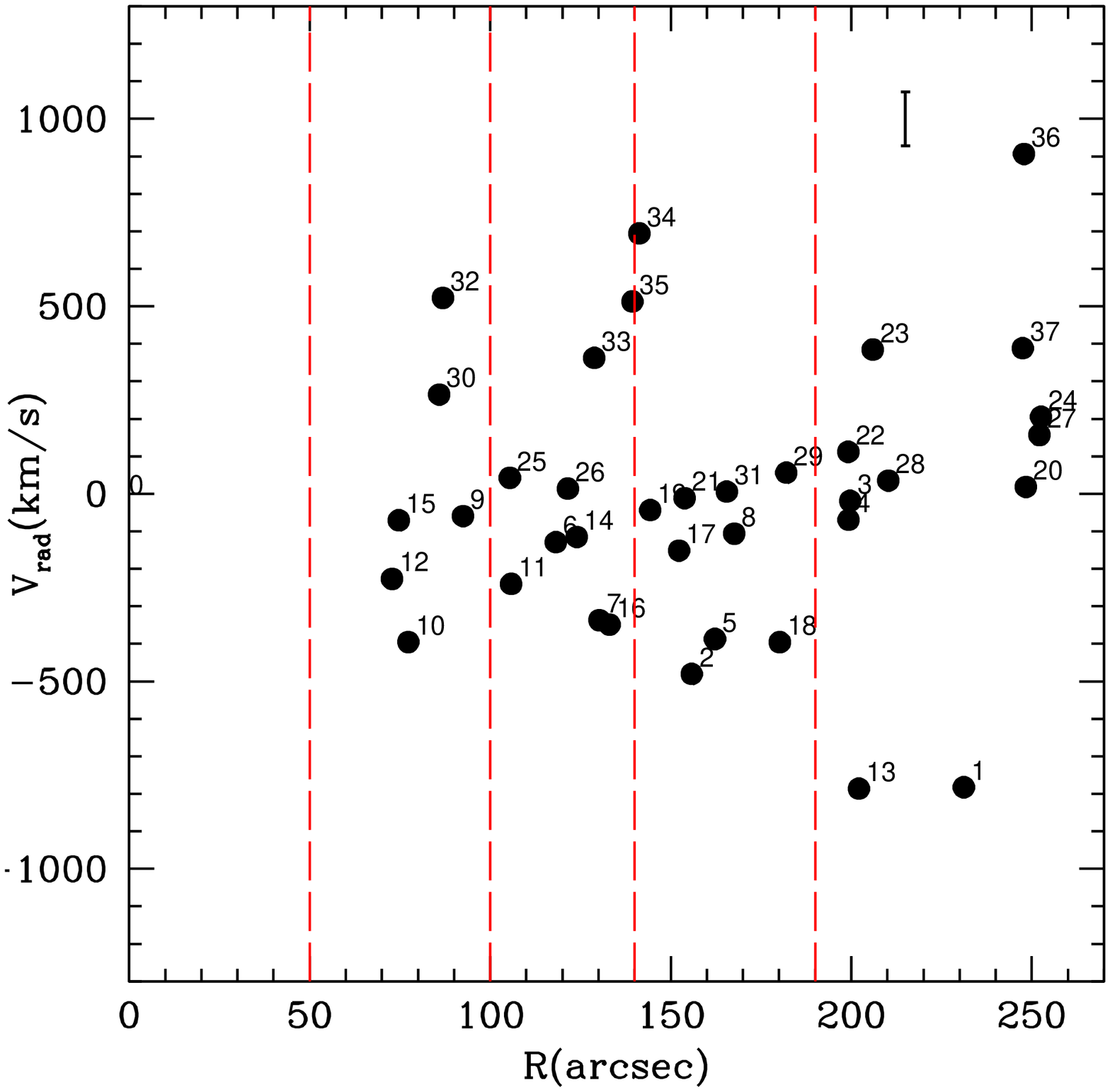,width=6.5cm,height=6.5cm}
\caption{\footnotesize  Left panel: radial velocities (open circles), which are selected along a 100$''$ wide strip about the Z1 axis, are plotted 
against distance along this axis. Full circles are the mean velocities
in each bin, while radial velocities from integrated light data  
(D'Onofrio et al. \cite{dono95}; Graham et al. \cite{grah98}; Saglia et al. \cite{sagl00}) are indicated with open symbols (triangles, squares, pentagons). 
The solid line is the bilinear fit and the dot-dashed line represents the flat curve fit to the data along Z1. Vertical dashed lines indicate the bin size
adopted for the kinematical estimates. Right panel: radial velocities (with the related PN ID numbers) 
are plotted as a function of the projected distance from the center, $R=\sqrt{X^2+Y^2}$. 
Vertical dashed lines indicate the width of radial annuli adopted in the 
kinematical analysis.}
\label{vradis}
\end{figure*}

Using Montecarlo simulations in spherical systems, NAFC showed that, even under SNS regime, simple fitting functions provide reliable estimates of the position angle, $\phi_\mathrm{Z1}$, of the kinematical major-axis, $Z1$. As in Arnaboldi et al. (\cite{arn98}) both a bilinear (BF) and a flat-curve (FC) function (Table 1) were interpolated to the radial velocity field, $V_\mathrm{rad}(X,Y)$, of the 37 NGC~1399 PNe from Arnaboldi et al. (\cite{arn94}), shown in Fig. \ref{field}. 
Clearly, the trend of the rotation curve, $V_\mathrm{rot}(Z1)$, is not necessarily matched by these functions, but NAFC proved that some information can be recovered by plain binning of the velocity data. In practice, they suggest to average the velocities of all the stars falling within a given range of $\vert Z1\vert$ and within a fixed distance, $dZ1$, from the Z1 axis; the bins are chosen in order to secure subsamples with 10 or more data points. As shown in NAFC this implies relative errors less than 30\% for both the velocity dispersion and rotation velocity. For NGC~1399 we have chosen three bins in $\vert Z1\vert$ and $dZ1=100''$. The binned radial velocities are shown in Fig. \ref{vradis} (left panel) together with the BF and FC projected along Z1 and $V_\mathrm{rot}$ in each bin.  

{\em Rotation -} The binned $V_\mathrm{rot}$ data (Table \ref{disp}) suggest that the PNe system rotates quite strongly inside $R\approx 140''$, while it does not in the outer regions. This behavior was not seen by Arnaboldi et al. (\cite{arn94}), who discussed only the results from the bilinear fit. 
Grillmair et al. (\cite{grill94}) and Minniti et al. (\cite{minn98}) do not report any rotation of the GCs system out to $\sim 8'$, while Kissler-Patig et al. (\cite{kisl99}) found evidence of rotation only at large radii ($V_\mathrm{max} $=153$\pm$93 km s$^{-1}$ for $R\ge5'$ along a P.A.=120$^\circ\pm$40$^\circ$). 
They attribute this feature to a tidal interaction or merging of cluster galaxies passing throughout the core of the Fornax Cluster, rather than to the sign of a relative high local angular momentum. We shall discuss this possibility in more details later. In any case, the lack of an inner rotation of the GC system supports its decoupling from the stellar component.\\
{\em Velocity dispersion -} In each Z1 bin, the projected velocity dispersion is the standard deviation of the sample (e.g. no-fit procedure, see NAFC), corrected for the measurement errors, $\sigma_\mathrm{p}=\sqrt{SD^2-\sigma^2_\mathrm{mea}}$, where $\sigma_\mathrm{mea}=$70 km s$^{-1}$ from Arnaboldi et al. (\cite{arn94}). Outside $R\approx 140''$, where no rotation is found, the velocity dispersion is obtained in radial annuli in order to retain all data (Fig. \ref{vradis}, right panel) and improve precision. The results are summarized in Table \ref{disp}. 
The velocity dispersion has a minimum at $R\sim 140''$, where $V_\mathrm{rot}$ drops, and increases outwards (see also Fig. \ref{dismod}). PN $\sigma_\mathrm{p}$ values are consistent with the integrated light's at small $R$ and with GCs' at large $R$.

This peculiar kinematical behavior possibly calls for a complex dynamical situation. 
In particular, the evidence that decoupled populations, such as PNe and GCs, exhibit the same ``heating'' in the outer regions appears peculiar. 
In the following we shall consider different scenarios for the dynamical status of the system, to be extensively analysed in the next section:

\underline{Case a).} The system, at equilibrium, has two decoupled dynamical components: an inner, rotating, cold component (possibly a disk) and an outer, non-rotating, hot component, and the transition region occurs at $R\sim140''$. In principle, we can expect some signatures of this multicomponent structure in the surface brightness profile. 
Improved photometric studies and a more detailed kinematics are needed for any significant treatments of this picture. The possibility of an inner extended disk as cold component seems, however, unlikely because such large disk sizes are not observed in ellipticals (Nieto et al. \cite{niet91}, van den Bosch et al. \cite{vdbor94}, Scorza et al. \cite{scor98}) or expected from multicomponent models (van den Bosch  \cite{vdbor98}): this scenario will not be included in the analysis which follows. 
 
\begin{table*}
\caption{Global kinematical quantities from the PNe radial velocity field. 
The first column gives the number of PNe considered (see discussion in the text). 
$V_\mathrm{sys}$ and $V_\mathrm{max}$ are in km s$^{-1}$ and $gradV$ are in km s$^{-1}$ arcsec$^{-1}$.}
\centering
\footnotesize
\begin{tabular}{c|ccc|cc}
\hline \hline
  \multicolumn {6}{c}{\normalsize{\bf PNe global kinematical quantities}}\\
\hline \hline 
\noalign{\smallskip}
 & \multicolumn {3}{|c}{\normalsize Bilinear Fit} & \multicolumn {2}{|c}{\normalsize Flat-Curve Fit}\\
 \hline
Sample & $V_\mathrm{sys}  $ & $\Phi_\mathrm{Z1} $ & gradV & $\Phi_\mathrm{Z1}$ & $V_\mathrm{max}$ \\
 37 PNe& 1502$\pm$60 & 143$^\circ\pm$15$^\circ$& 1.1$\pm$0.9 & 140$^\circ\pm$15$^\circ$ & 177$\pm$91\\
 34 PNe& 1521$\pm$49 & 134$^\circ\pm$12$^\circ$& 0.52$\pm$0.11 & 137$^\circ\pm$13$^\circ$ & 107$\pm$23\\
\hline\hline
\noalign{\smallskip}
\end{tabular}
\label{glqu}
\end{table*}

\underline{Case b).} In a more conservative way, if PNe are a system at equilibrium in the gravitational potential of NGC~1399, the rise of $\sigma_\mathrm{p}$ (i.e. of the pressure gradient) is needed to compensate the lack of rotation at large radii.  In this and the above picture, the external regions are fading into the cluster as we can infer from both the luminosity and velocity dispersion profiles which are consistent with the cluster profiles (Grillmair et al. \cite{grill94}, Arnaboldi et al. \cite{arn94}, Kissler-Patig et al. \cite{kisl99}). 
In the following dynamical analysis, we will refer to this as \underline{Model I}.

\underline{Case c).} An alternative picture is that the regions at radii $R>140''$ experienced some forms of interaction with the cluster. This possibility has been invoked in previous works on PNe and GCs kinematics and also in X-ray studies (Rangajan et al. \cite{rang95}; Ikebe et al. \cite{ike96}; Jones et al. \cite{jon97}; Paolillo et al. \cite{paol01}). In this case there might be objects which are no more bound to NGC~1399. 
\begin{figure*}
\centering
\epsfig{figure=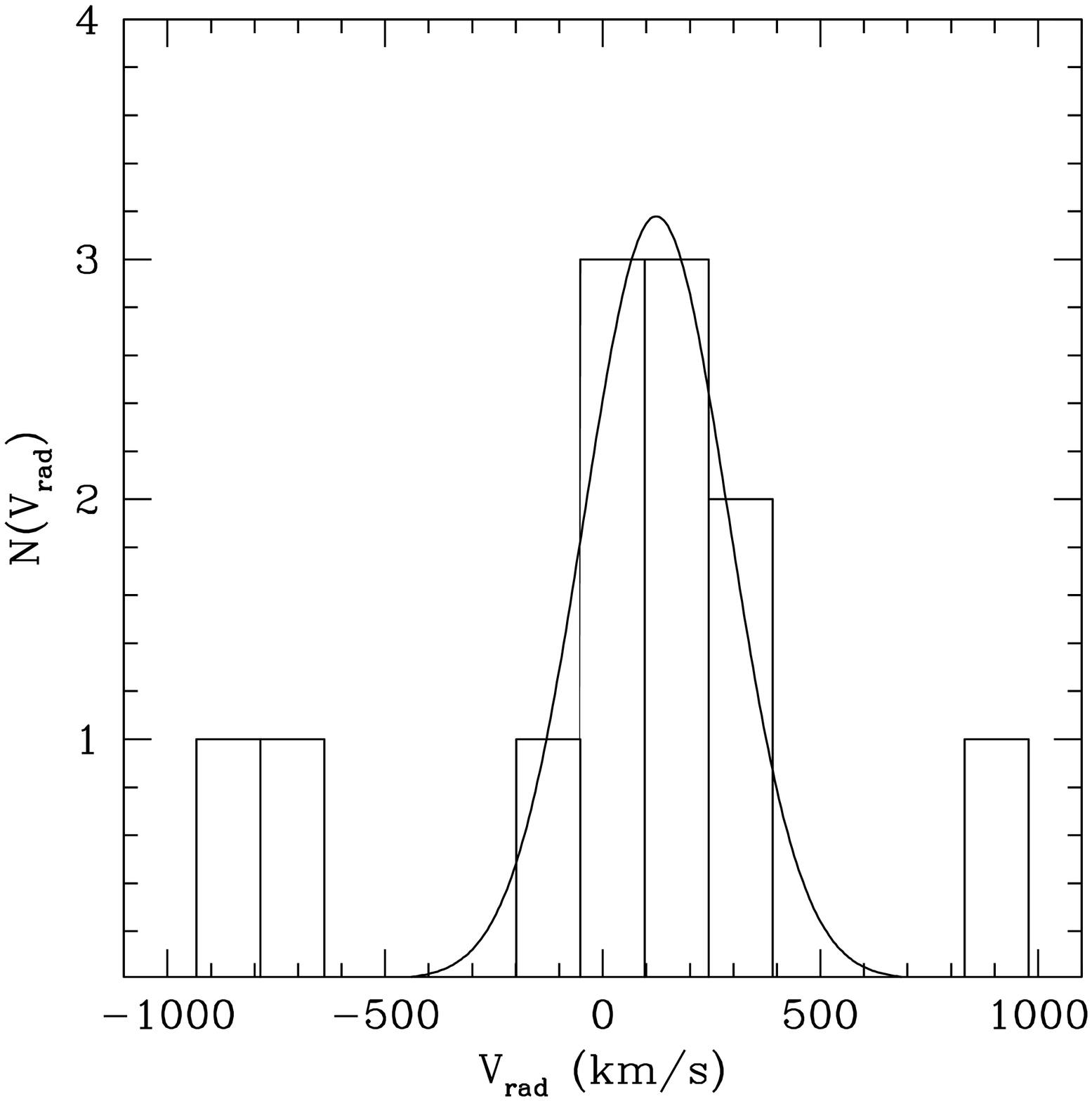,width=6cm,height=6cm}
\epsfig{figure=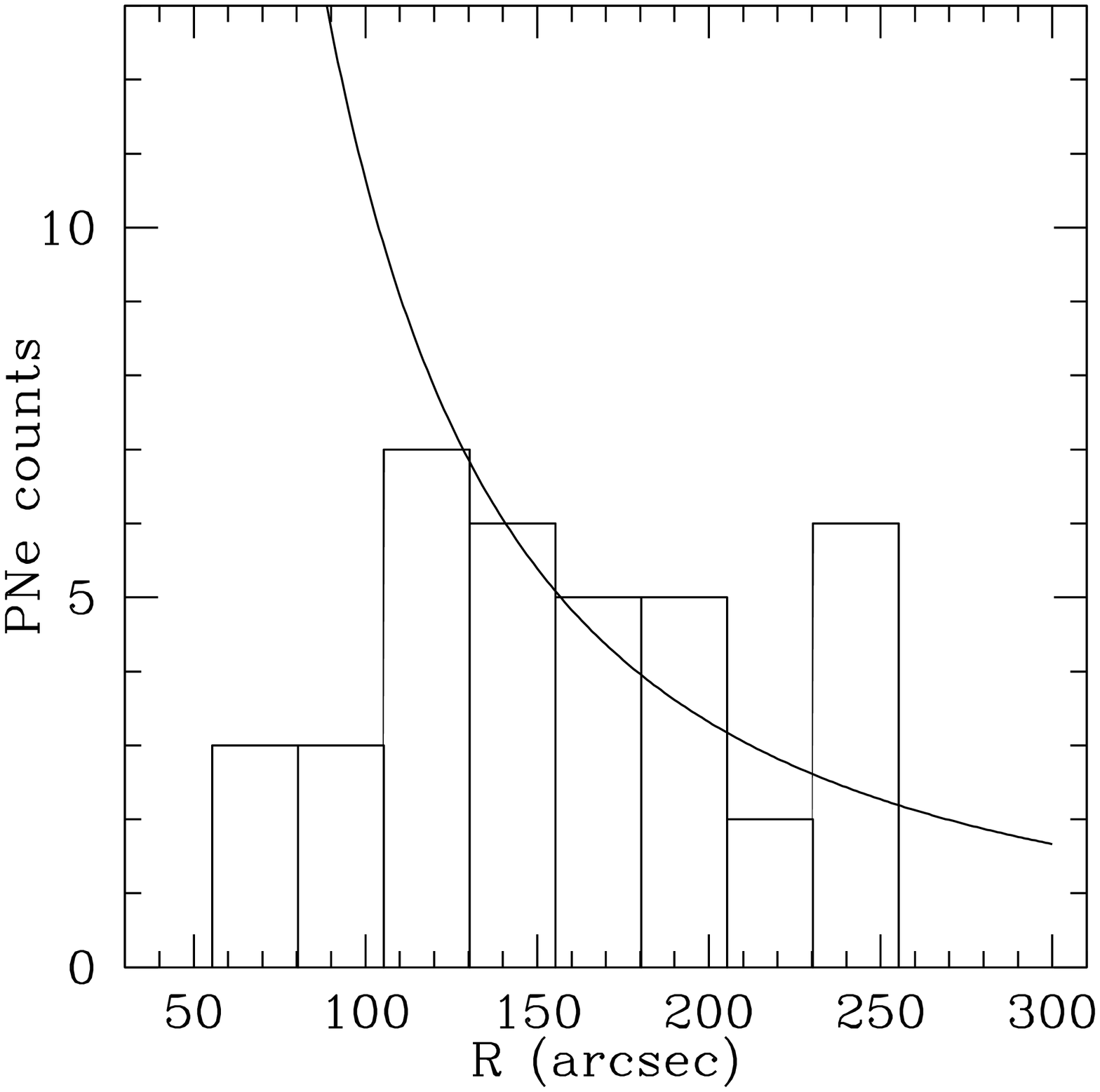,width=6cm,height=6cm}
\caption{\footnotesize Left panel: Distribution of the radial velocities in the last radial bin, $R_\mathrm{last}=225''$ (see Fig.\ref{vradis}). 
Three PNe (ID=1,13,36 in Fig. \ref{vradis}) are inconsistent with the Gaussian-like 
behavior of the remaining objects. Right panel: PNe radial number density profile. PNe follow the 
surface brightness profile out to 200$''$: at larger radii there is a
possible excess of PNe.}
\label{hist}
\end{figure*}
There are three outliers in the distribution of $V_\mathrm{rad}$ for the last radial bin (Fig. \ref{hist}). Their large velocities are inconsistent with the Gaussian-like behavior expected for the radial velocity distributions based on small statistical samples\footnote{ The observed distribution of radial velocity profiles in early type galaxies have typical deviation of 10\% from a Gaussian-like behavior (Winsall \& Freeman \cite{wifr93}; Bender et al. \cite{bend94}). In order to appreciate these deviations, more than few hundreds of radial velocities are needed (Merritt \cite{merr97}).}. Moreover they are responsible for the number overdensity in the outer regions of the galaxy (see Fig. \ref{hist}). 
Were NGC~1399 a relaxed system, these objects should not be included when computing the mass at equilibrium (\underline{Model II}, hereafter). 
The global kinematical quantities computed after discarding PN ID=1, 13 and 36 
(Fig. \ref{vradis}), are listed in Table \ref{glqu}, and the rotation 
velocity and velocity dispersion values in Table \ref{disp}.

\begin{table*}
\caption{Projected kinematical quantities (given in km s$^{-1}$). Errors on the rotation velocity and velocity dispersion are $SD/\sqrt{N}$ and $SD/[\sqrt{2N}(1-\sigma_\mathrm{mea}^2/SD ^2)^{1/2}]$ respectively. The latter is obtained by propagating the measurement errors. The code in the last column is: NF=no-fit procedure, BF=bilinear-fit procedure, R=radial annuli, Z1=Z1 bins. For an extensive description of the procedures, see Napolitano et al. (\cite{napo01}). $^{(*)}$Standard deviation of the sample of the residual field $\Delta V=V_\mathrm{rad}-BF$ (see NAFC).}
\centering
\begin{tabular}{ccc|ccc|c}
\hline \hline 
  \multicolumn {7}{c}{\normalsize{\bf PNe projected kinematical quantities}}\\
\hline \hline 
\noalign{\smallskip}
  \multicolumn {3}{c|}{\normalsize 37 PNe}& \multicolumn {3}{c|}{\normalsize 34 PNe}& \\
 \hline
Distance & $V_\mathrm{rot}$ & $\sigma_\mathrm{p}$ &Distance& $V_\mathrm{rot}$ & $\sigma_\mathrm{p}$ &Procedure\\
 39$''$  & 16$\pm$121 & 297$\pm$108& 28$''$  &  -3$\pm$156  & 304$\pm$110 & NF(Z1)\\
 101$''$ & 251$\pm$77 & 244$\pm$57 & 101$''$ &  225$\pm$75  & 262$\pm$58 & NF(Z1)\\
 116$''$ &            & 260$\pm$70 & 116$''$ &  	    & 257$\pm$60 & BF(R)$^{(*)}$\\
 156$''$ & 	         & 334$\pm$77 &  157$''$ &         & 188$\pm$47 & NF(R)\\
 182$''$ & 8$\pm$107  & 360$\pm$90 & 173$''$ &  -80$\pm$70  & 195$\pm$49 & NF(Z1)\\
 225$''$ & 	      & 460$\pm$96 & 224$''$ &         & 150$\pm$39 & NF(R)\\
\hline
\noalign{\smallskip}
\end{tabular}
\label{disp}
\end{table*}

\underline{Case d).} Finally, we assume that the outer halo of NGC~1399 is not at equilibrium (\underline{Model III}, hereafter). This picture is treated separately in the next Section.

\section{Dynamical analysis}
By combining PN and GC data with the inner stellar kinematics it is possible to infer the mass distribution out to the halo regions of NGC~1399\footnote{This is strictly true for PNe because they share the same dynamics than the inner stellar population. As stressed by Saglia et al. (\cite{sagl00}), GCs number density profile as function of radius follow a power law  with a slope which is 0.28dex shallower (Forbes et al. \cite{forb98}). Therefore, we expect GCs to have a velocity dispersion $\approx$ 1.08 larger than stars and PNe; when they are used together with the stellar kinematics the velocity dispersion has to be corrected for this factor.}.
As in previous work with larger samples (Hui et al. \cite{hui95}), one solves the Jeans equations in the equatorial plane and obtains the mass distribution in some analytical form. NGC1399 is a nearly round (E0-E1) galaxy. Its dynamics depends on the ellipticity of the potential, which is  much rounder than its density distribution ($\epsilon_{\Phi}\sim\epsilon_{\rho}/3$; Binney \& Tremaine \cite{bintr}). Therefore, spherical symmetry is adopted for this system.

For sake of simplicity, we assume isotropy for the velocity ellipsoid in the Jeans equations, accordingly with previous works based on discrete radial velocity fields in different early-type systems (Grillmair et al. \cite{grill94}, Hui et al. \cite{hui95}, Arnaboldi et al. \cite{arn98}, Kissler-Patig et al. \cite{kisl99}). Indeed, isotropy is not excluded by previous dynamical studies, based on integrated light data: i) Bicknell et al. (\cite{bick89}) found a variable anisotropy parameter within $100''$ around a central null value for a constant mass-to-light ratio, $M/L_\mathrm{B}=14$; ii) Saglia et al. (\cite{sagl00}) found that a wide range of anisotropy parameter, including the isotropic case, are compatible with the dispersion profiles from PNe and GCs. 
On the equatorial plane ($z=0$), the isotropic axisymmetric Jeans equation 
\begin{equation}
\frac{\partial \rho(R,0) \sigma_\mathrm{R}^2(R,0)}{\partial R} - \frac{\rho(R,0)}{R} 
v_\mathrm{rot}^2(R)
= - \rho(R,0) \frac{\partial \Phi(R,0)}{\partial R}
\label{axJ1}
\end{equation}
is equivalent to the Isotropic Radial Jeans Equation (IRJE) in spherical coordinates. Using $\rho(R,0)=\rho(r)$ from the light profile with a fixed M/L, $V_\mathrm{rot}$ from the observations and adopting an analytical form for the potential, Eq. (\ref{axJ1}) can be solved and the mass distribution derived from the velocity dispersion data. In more details:\\
1) under spherical symmetry, we compute $\rho(r)$ by deprojecting the Killeen \& Bicknell's (\cite{kill88}) surface brightness profile
\begin{equation}
\Sigma_\mathrm{B}(R)=\Sigma_0\left(1+\frac{R}{R_\mathrm{s}}\right)^{-\alpha}
\label{SBKB}
\end{equation}
with $\Sigma_\mathrm{0}=16.73\pm 0.05$ mag$_\mathrm{B}$ arcsec$^{-2}$, $R_\mathrm{s}=3.0''\pm 0.2''$ and $\alpha=1.72\pm0.05$ and the mass-to-light ratio, $M/L_\mathrm{B}=12M_{\odot}/L_{\odot}$ (we adjust the M/L from the same authors to our adopted distance). This profile is consistent with the PNe number density distribution (see Fig. \ref{hist}).  
Saglia et al. (\cite{sagl00}) performed an accurate deprojection of the light 
distribution and a dynamical modeling for NGC~1399, obtaining a $M/L_\mathrm{B}=10M_{\odot}/L_{\odot}$ for the luminous component inside $R=60''$,
where the influence of the dark halo starts. Since they do not provide us with the stellar density distribution in an analytical form, we prefer to use the analytical expression from Eq. (\ref{SBKB}) (as was done by Kissler-Patig et al. \cite{kisl99}, for instance).\\
2) Due to the lack of detailed data, a simplified model of $V_\mathrm{rot}(R)$ is assumed: we have a linear regime out to $R\approx140''$, then $V_\mathrm{rot}$ drops to zero. 
This artificial model is deprojected assuming circular orbits and no correction is adopted for inclination (the reason is that the peak velocity is already large enough to accommodate a high inclination).\\  
3) We consider a dark mass contribution to the potential by a simple pseudo-isothermal sphere: 
\begin{equation}
\rho_\mathrm{d}=\frac{\sigma_\mathrm{d}^2}{2\pi G (r^2+r_\mathrm{d}^2)}
\label{rhod}
\end{equation}
where $\sigma_\mathrm{d}$ and $r_\mathrm{d}$ are free parameters. In the IRJE:  
\begin{eqnarray}
\frac{\partial \Phi (r)}{\partial r}= G \frac{M_\mathrm{lum}(r)+M_\mathrm{d}(r)}{r^2}= & \nonumber \\ 
=\frac{4\pi G}{r^2} \int_0^r (\gamma j(r')+ \rho_\mathrm{d}(r')) r'^2 dr' &
\end{eqnarray}
where $\gamma=M/L_\mathrm{B}$ and $j(r)$ is the deprojected luminosity profile;

4) $\sigma_\mathrm{d}$ and $r_\mathrm{d}$ are assigned by matching the velocity dispersion
solution of the IRJE, once projected along the line-of-sight, to the observed velocity dispersion estimates by minimizing the $\chi^2$.\\
This procedure implicitly assumes that the system is at equilibrium.
\subsection{The equilibrium analysis: mass distribution}
{\em Model I}: We consider the complete sample of 37 PNe plus the GCs data from Kissler-Patig et al. (\cite{kisl99}).
\begin{figure*}
\centering
\epsfig{figure=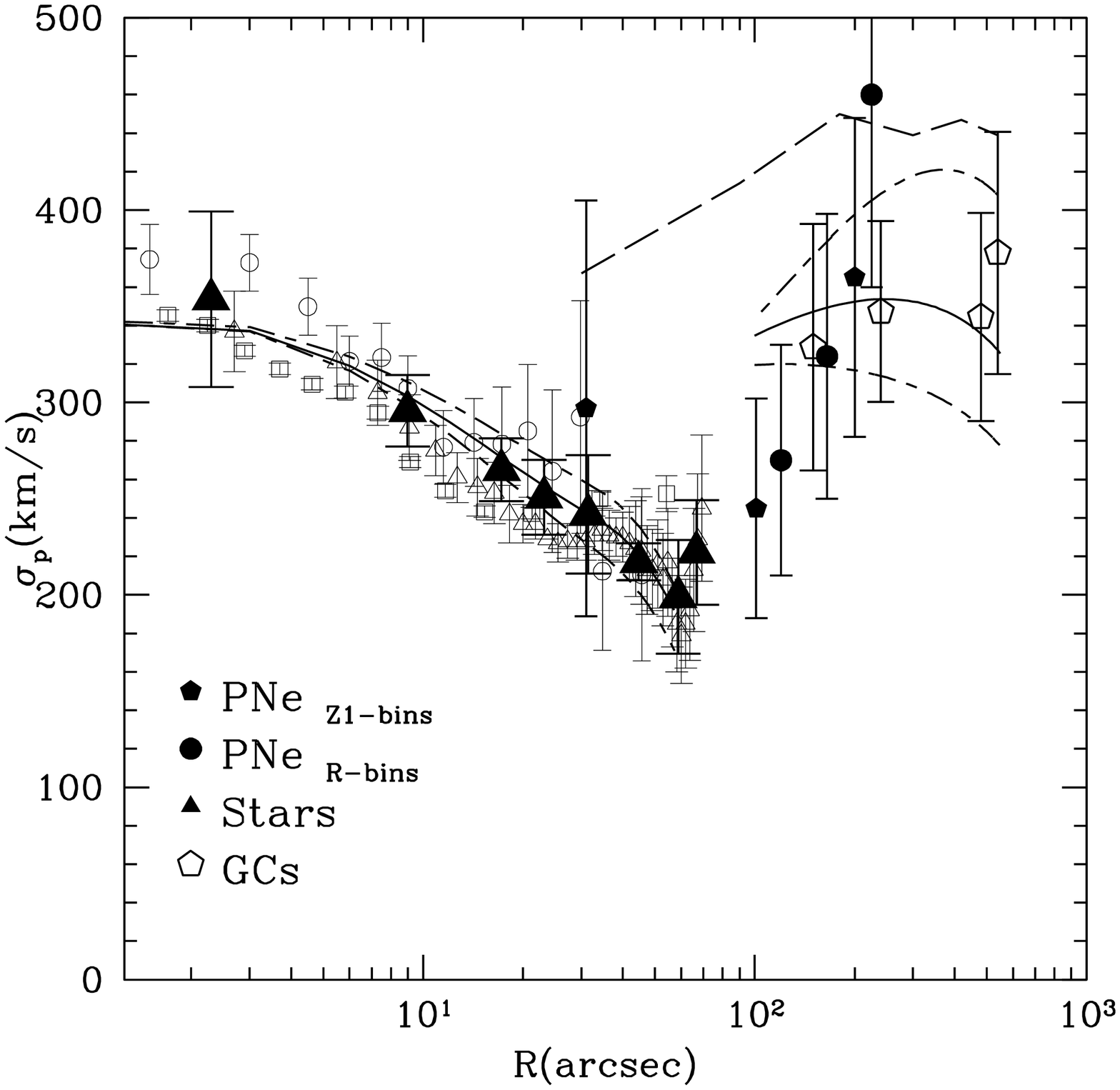,width=6cm,height=6cm}
\epsfig{figure=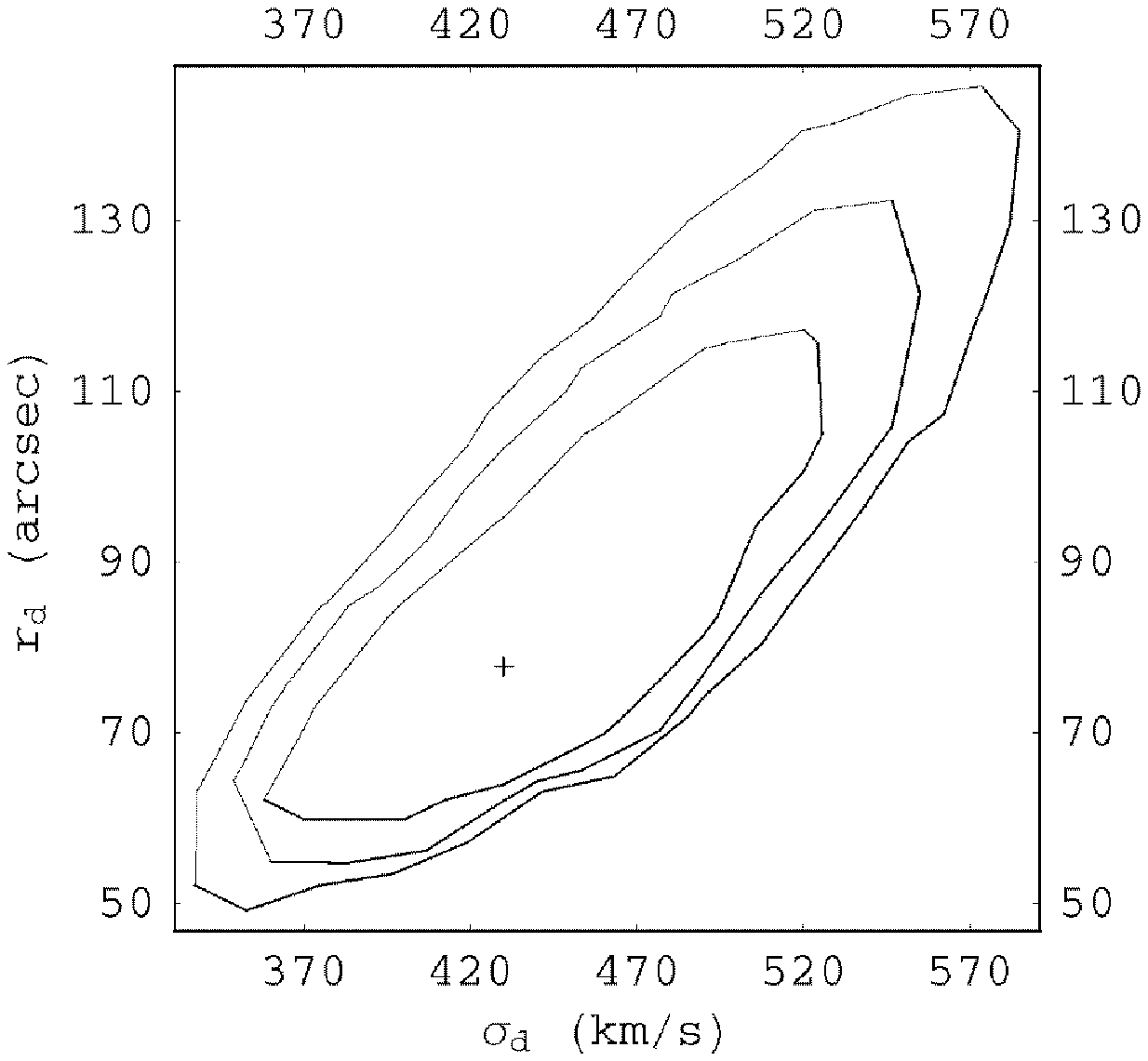,width=6cm,height=6cm}
\epsfig{figure=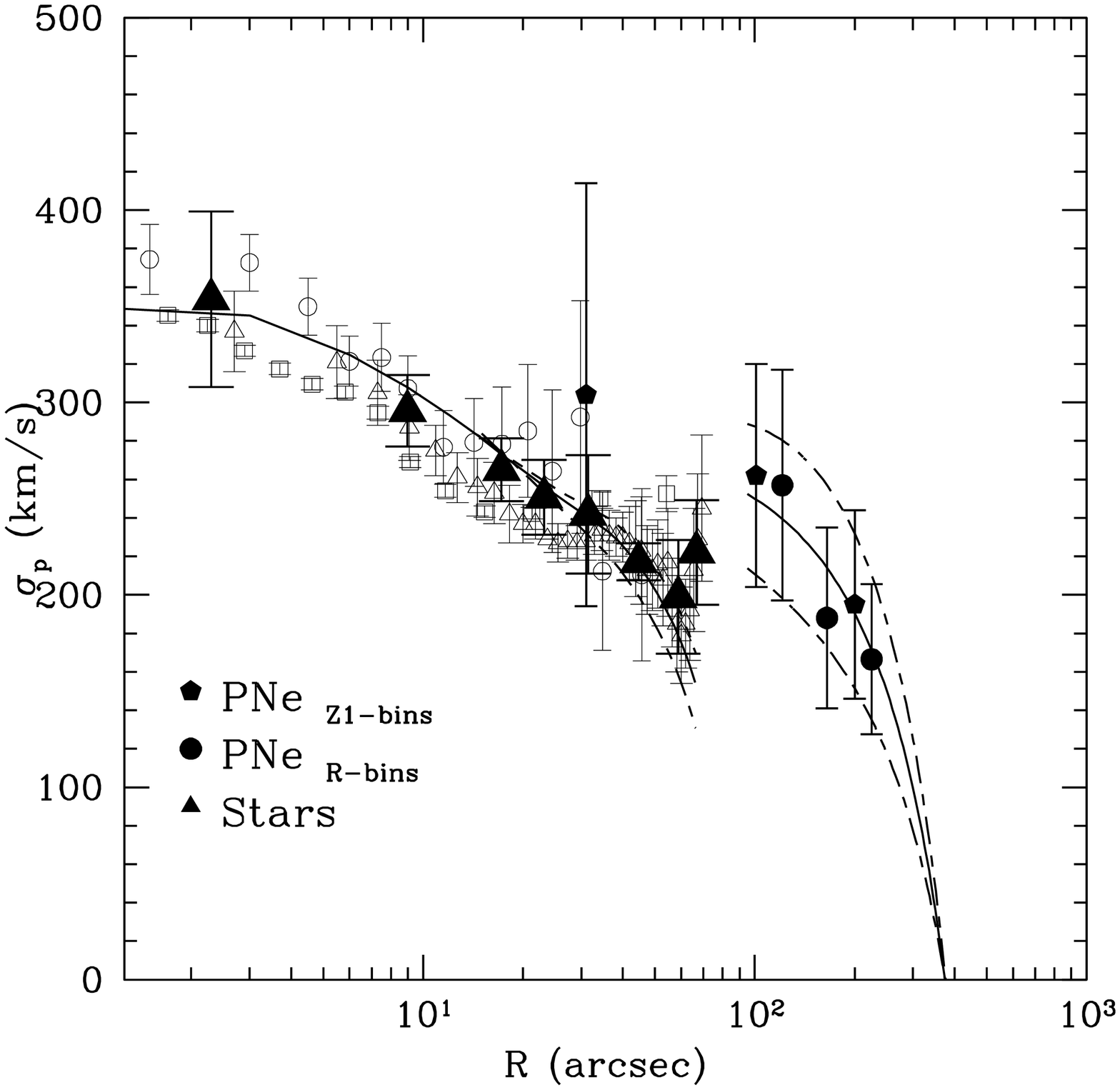,width=6cm,height=6cm}
\epsfig{figure=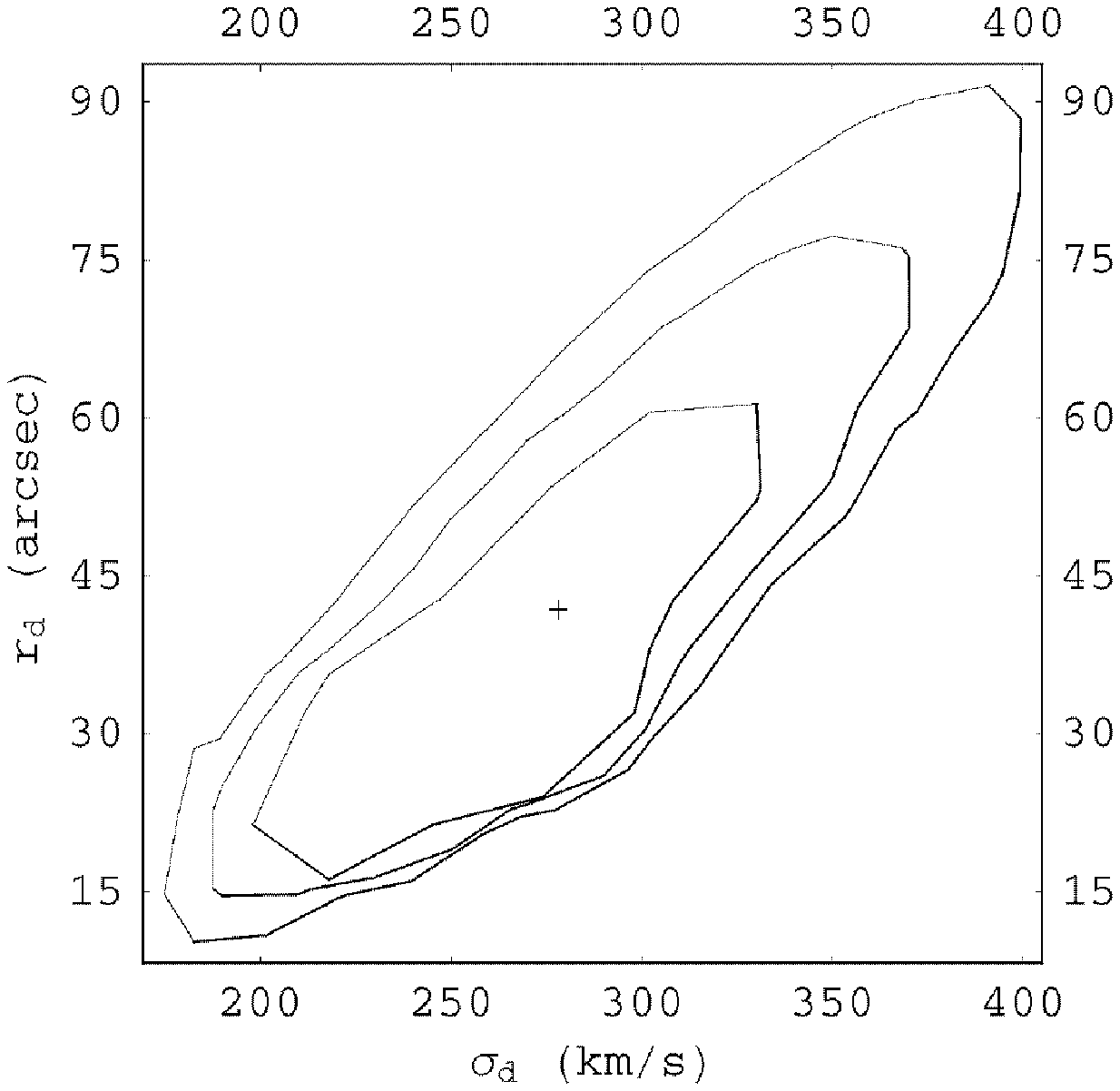,width=6cm,height=6cm}
\caption{\footnotesize Velocity dispersion profiles observed and modeled and the confidence regions for the model parameters. 
Upper left box: the PNe estimates in Z1 bins (filled pentagons) and radial annuli 
(filled circles) are plotted with the mean stellar kinematics in spatial bins (filled triangles) obtained from 
different authors (plotted with different symbols) and GCs estimates 
(open pentagons, corrected for the factor 1.08, see discussion in the text). 
Under equilibrium hypothesis (Model I), the solid line is the best-fit 
($\sigma_\mathrm{d}$=430 km s$^{-1}$, $r_\mathrm{d}$=78$''$, 98\% significance level) to the data, 
the dashed line is the velocity dispersion profile derived 
from the temperature of the X-ray gas (Jones et al. \cite{jon97}). 
Dash-dotted lines correspond to the extreme parameter values 
in the 68\% confidence region. Upper right box: we show
the confidence regions at 68\%, 90\%, 95\% in the parameter space. 
Lower left box: velocity dispersion estimates for the 34 PNe sample. Solid and dash-dotted lines as above: here they are computed for the Model II as best-fit to the data ($\sigma_\mathrm{d}$=280 km s$^{-1}$, $r_\mathrm{d}$=44$''$, 98\% s.l.). In this case no GCs data are used. Lower right box: the 68\%, 90\%, 95\% confidence levels for Model II. 
}
\label{dismod}
\end{figure*} 
The fit to the velocity dispersion profile is shown in Fig. \ref{dismod} 
with the confidence regions for the best-fit parameters ($\sigma_\mathrm{d}$=430 km s$^{-1}$, $r_\mathrm{d}$=78$''$). The mass distribution is shown in Fig. \ref{mod}: it gives  $M/L_\mathrm{B}=56^{+15}_{-13}M_{\odot}/L_{\odot}$ within 400$''$, where the quoted errors are related to the extreme values of $\sigma_\mathrm{d}$ and $r_\mathrm{d}$ in 68\% confidence region and produce the dot-dashed dispersion models in Fig. \ref{dismod}.\\
\\
{\em Model II}: Here only 34 PNe (the three PNe (ID = 1, 13, 36) were excluded) are considered to be at equilibrium. Fig. \ref{dismod} shows the fit to the velocity dispersion estimates with the related confidence regions in parameter space. The best-fit parameter are $\sigma_\mathrm{d}$=280 km s$^{-1}$ and $r_\mathrm{d}$=44$''$. In this case, we introduced a tidal radius, $R_\mathrm{t}$=360$''$, where the stellar distribution 
is truncated, which has improved the fit to the data. This possibly indicates the most external radius of NGC~1399, while the cluster potential dominates at larger radii. The derived mass distribution is also shown in Fig. \ref{mod}: in this case we obtain $M/L_\mathrm{B}=33^{+16}_{-14}M_{\odot}/L_{\odot}$ within 400$''$.\\

Despite the oversimplified approximation for the rotation curve, the above analysis provides the mass distribution needed to match the velocity dispersion profile in the two extreme rotation regimes (i.e. rigid rotation within R$\approx$140$''$, null rotation outside). 
The link between the two regions calls for a more 
detailed kinematics at the intermediate radii. 
The mass distribution from our analysis is in perfect agreement with estimates from Saglia et al. (\cite{sagl00}) who found M=1.2-2.5$\times$10$^{12} M_{\odot}$ within R=400$''$.
\begin{figure*}
\centering
\epsfig{figure=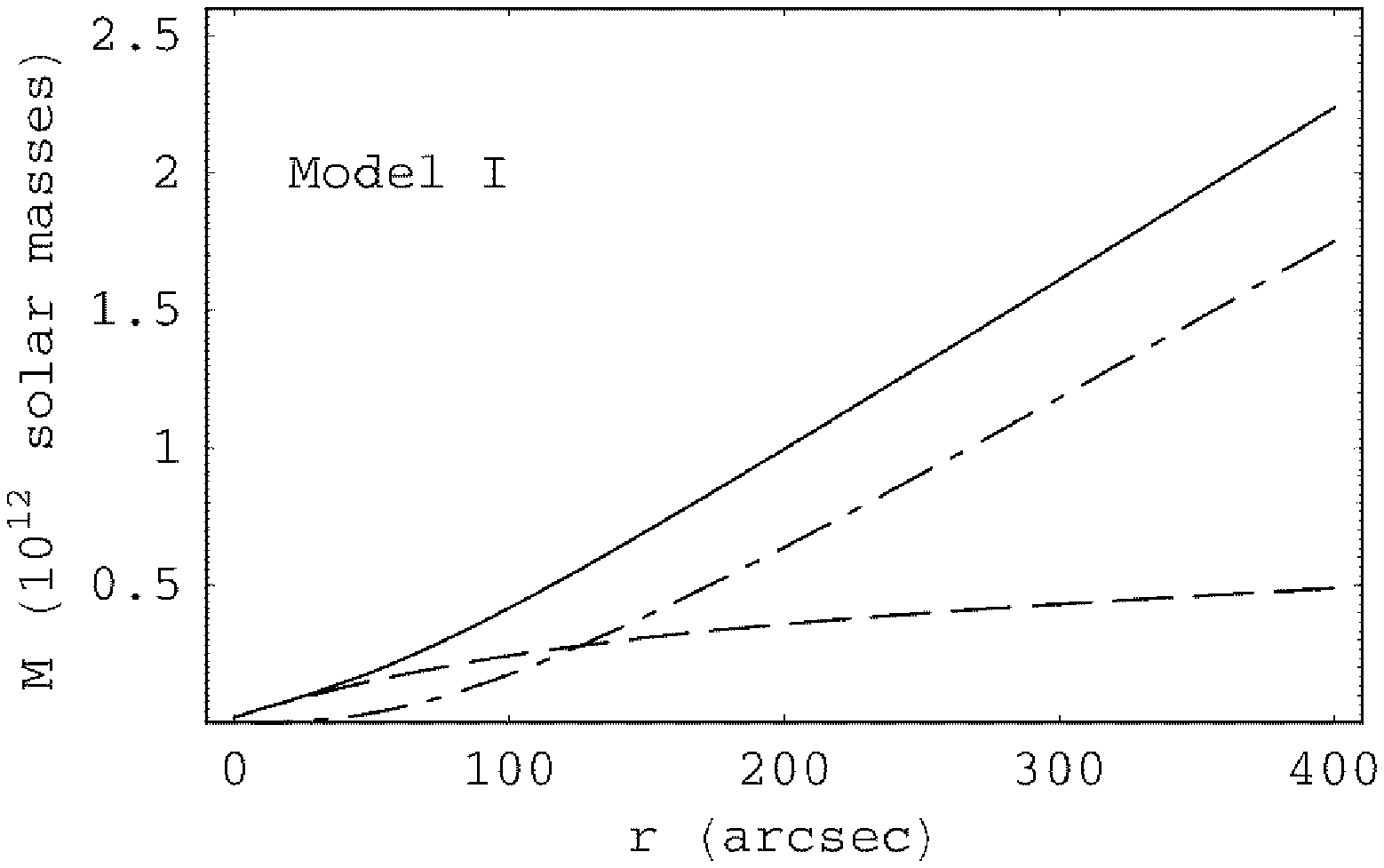,width=6.7cm,height=6.7cm,angle=-90}
\epsfig{figure=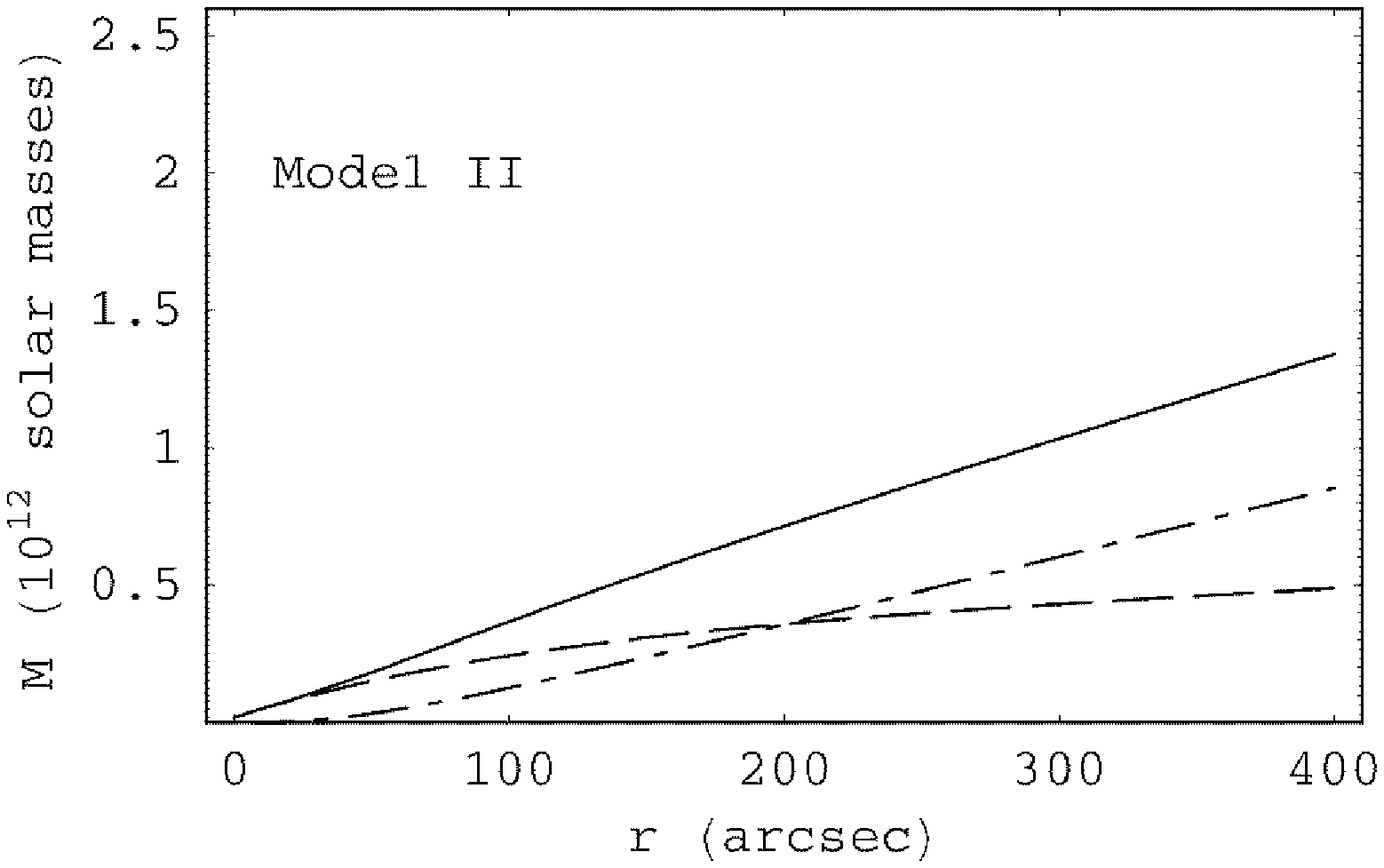,width=6.7cm,height=6.7cm,angle=-90}
\caption{\footnotesize The modeled mass distributions for 
Model I (left) and Model II (right). 
Dashed line is the luminous mass using Killeen \& Bicknell (\cite{kill88}) luminosity 
profile with $M/L_\mathrm{B}=12M_{\odot}/L_{\odot}$, 
dash-dotted line is the dark halo mass from Eq. (\ref{mod}) 
where the best-fit parameters were adopted, 
and the solid line is the total mass, $M_\mathrm{tot}$.}
\label{mod}
\end{figure*}


\subsection{The non-equilibrium scenario for NGC~1399}
The peculiar kinematics and the temperature profile of the hot gas from the X-ray observations may suggest that NGC~1399 is not a relaxed system. 
Minniti et al. (\cite{minn98}) and Kissler-Patig et al. (\cite{kisl99}) use the metallicity distribution and specific density of GCs to claim evidence for a recent interaction with the nearby companion NGC~1404.
If such interaction occurred, the most spectacular effect might be the strong heating of the intergalactic gas.
\begin{table}
\caption{Projected kinematical quantities (units are km s$^{-1}$) for the PNe subsample within 140$''$.}
\centering
\begin{tabular}{cccc}
\hline \hline
  \multicolumn {4}{c}{\normalsize{\bf PNe projected kinematical quantities}}\\
\hline \hline 
\noalign{\smallskip}
 & \multicolumn {2}{c}{\normalsize Model III}& \\
 \hline
Distance& $V_\mathrm{rot}$ & $\sigma_\mathrm{p}$ &Procedure \\
 33  & 109$\pm$104 & 221$\pm$73 & NF(Z1)\\
 88  & 	      & 244$\pm$63 & BF(R)\\
 94  & 200$\pm$72 & 192$\pm$51 & NF(Z1)\\ 
 126 & 	      & 146$\pm$40 & BF(R)\\
\hline
\noalign{\smallskip}
\end{tabular}
\label{disnoeq}
\end{table} 
The hot gas in NGC~1399 has a temperature which, once converted into a velocity dispersion, seems systematically larger by $\sim$30 km s$^{-1}$ than that of the stellar population (see the long-dashed curve in Fig. \ref{dismod}). Moreover, the decline of the rotation curve of NGC~1399 is similar to those observed in spiral systems and reproduced by single and/or multiple galaxy encounter (Kauffmann \cite{kauf99}; Salo \& Laurikainen \cite{salau00}).\\ 
We shall then explore a model where NGC~1399 is out of equilibrium, and the stars in the outer halo are re-arranging themselves after a flyby encounter with NGC~1404. NGC~1399 and NGC~1404 have a small projected distance, $b=9'$, and a large relative radial velocity, $V=522$ km s$^{-1}$ ($V_{1399}=1425\pm 5$, Graham et al. \cite{grah98}, $V_{1404}=1947\pm 5$, D'Onofrio et al. \cite{dono95}). 
The simplest approach is to use the general scheme of the {\em impulse approximation} (Binney \& Tremaine \cite{bintr}), where NGC~1399 is the perturbed system and NGC~1404 is the perturber.
In this approach, the encounter does not modify the potential of the perturbed system, and the kinetic energy in the inner regions, while the outer regions ($R \ge 140''$) experienced 1) an energy injection, 2) the disruption of the streaming motions, i.e. of the angular momentum (Sugerman et al. \cite{suka00}), both re-distributed in the random motions, and 3) no mass loss (we will discuss this assumption later). Following this scheme, the post-encounter kinetic energy of the perturbed system, $E_\mathrm{fin}$, can be written as
\begin{equation}
E_\mathrm{fin}=E_\mathrm{in} +\Delta E
\label{efin}
\end{equation} 
where $E_\mathrm{in}$ is the kinetic energy before 
the encounter, that is at equilibrium, and $\Delta E$ is the variation in the kinetic energy induced by the encounter.
In Eq. (\ref{efin}), $E_\mathrm{fin}$ is associated to the observed random motions only, which are the result of the energy injection and the disruption of the streaming motions; it can be computed as $E_\mathrm{fin}=3/2 M \sigma_\mathrm{obs}^2$. Let us estimate $E_\mathrm{in}$ and $\Delta E$.\\   
\begin{figure}
\centering
\vspace{-1cm}
\epsfig{figure=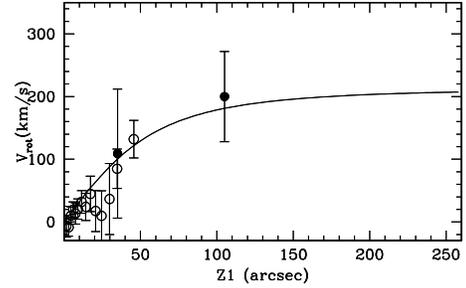,width=6.3cm,height=6.3cm}
\epsfig{figure=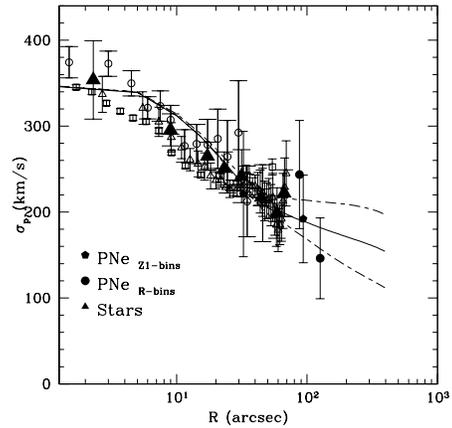,width=6cm,height=6cm}
\caption{\footnotesize Model III. Top panel: rotation velocities from stars (open circles) and PNe (full circles) 
with the modeled rotation curve from Eq. (\ref{roneq}). Bottom panel: velocity dispersion 
estimates and the best-fit to the data within $140''$. Symbols have the same meaning as in Fig. \ref{dismod}. 
}
\label{curnoeq}
\end{figure}
{\em The equilibrium before the encounter - } $E_\mathrm{in}$ writes as 
\begin{equation}
E_\mathrm{in}=1/2 M V_\mathrm{rot}^2 + 3/2 M \sigma_\mathrm{eq}^2,
\end{equation}
where $V_\mathrm{rot}$ and $\sigma_\mathrm{eq}$ are derived from the kinematical data of both the integrated light and PNe in the range of radii where the system is assumed to be still at equilibrium ($R < 140''$). In particular, for the PNe system we obtain $V_\mathrm{sys}=1465\pm73$ km s$^{-1}$, $\Phi_\mathrm{Z1}=131^\circ \pm 7^\circ$, $gradV= 2.1\pm0.3$ km s$^{-1}$ arcsec $^{-1}$ (BF). Here we emphasize that the $V_\mathrm{sys}$ estimate, from the inner PNe subsample, is in good agreement with the stellar light data. The estimates in Table \ref{glqu} show a positive residual in the outer regions, i.e. in the same direction of the assumed encounter between NGC 1399 and NGC 1404. This effect is expected in the impulse approximation (Aguilar \& White \cite{agwi85}), and supports our assumption about the inner equilibrium.

$V_\mathrm{rot}$ and $\sigma_\mathrm{p}$ are listed in Table \ref{disnoeq} and shown in Fig. \ref{curnoeq}. In the left panel $V_\mathrm{rot}$ is interpolated with a flat rotation curve of the form\footnote{This functional form allows a good description for an extended flat rotational curve (Hui et al. \cite{hui95}) as observed in different ellipticals (NGC 5128: Hui et al. \cite{hui95}; GC 3115: Capaccioli et al. \cite{capa93}; Emsellem et al. \cite{emsel99}; NGC1316: Arnaboldi et al. \cite{arn98}).}
\begin{equation}
V_\mathrm{rot}(R)=\frac{V_0 R}{\sqrt{R^2+R_0^2}}
\label{roneq}
\end{equation}
where $V_0=214$ km s$^{-1}$ and $R_0=67''$. This is used in the IRJE to fit $\sigma_\mathrm{p}$ as in Sect.~3 and obtain the modeled $\sigma_\mathrm{eq}$. 
\begin{figure}
\centering
\epsfig{figure=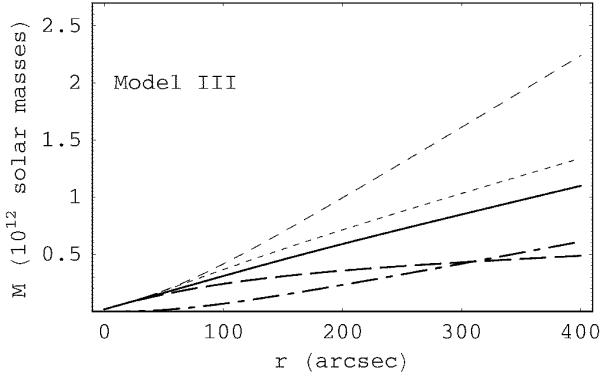,width=8cm,height=8cm,angle=-90}
\vspace{-1cm}
\caption{\footnotesize Model III. The total mass distribution (heavier solid line) for Model III
is compared with mass distributions from Model I (short-dashed line) and II (dotted line). Heavier long-dashed line and dot-dashed have the same meaning than as in Fig. \ref{mod}. The contribution of the dark halo 
becomes important for $R > 60''$ in all models.}
\label{noeqmo}
\end{figure}
The `` best-fit'' is shown in Fig. \ref{curnoeq} (right panel): the parameter are $\sigma_\mathrm{d}$=230$^{+62}_{-74}$ km s$^{-1}$, $r_{\mathrm{d}}=67^{+16} _{-28}$ arcsecs ($P(\tilde{\chi}^2>\tilde{\chi}_0^2)=$97\%, quoted errors are 68\% c.l.) and the mass distribution is in Fig. \ref{noeqmo}. The mass-to-light ratio is
$M/L_\mathrm{B}=26^{+8}_{-6} M_{\odot}/L_{\odot}$ within $400''$, where the errors refer to the extreme values of the 68\% confidence region.\\     
{\em The injected energy $\Delta E$ - }Following Binney \& Tremaine (\cite{bintr}), we estimate the injected energy using the {\em tidal approximation} equation
\begin{equation}
\Delta E=\frac{4G^2M_2^2M_1}{3b^4V^2}\overline{r^2}
\label{delE}
\end{equation}
where $M_1$ and $M_2$ are the total masses of NGC~1399 and NGC~1404 respectively, 
$b$ is the impact parameter, $V$ is the relative velocity, 
$\overline{r^2}$ is the squared mean radius of the perturbed system and $G$ is the gravitational constant. 
We initially assume a tangential encounter, where $b=9'$ is the projected
separation between NGC~1399 and NGC~1404, and the relative velocity is then given
by the difference of the object systemic velocities, $V=522$ km s$^{-1}$. 
In Eq. (\ref{delE}), we adopt $M_1=1.7\times10^{12} M_\odot$ and $(\overline{r^2})^{1/2}=40$ kpc from our mass model.
$M_2$ is computed from NGC~1404 total luminosity 
$L_\mathrm{B}=1.84\times10^{10}L_{\odot}$ (de Vaucouleurs \cite{devau91} and D=17 Mpc) and a constant $M/L_\mathrm{B}$=17 $M_\odot/L_\odot$,
similar to NGC~1399. Finally we obtain $\Delta E=9\times10^{15}M_\odot$ km$^2$ s$^{-2}$, which is $\sim$7\% of the total energy.\\ 
Following Aguilar \& White (\cite{agwi85}), this is consistent with the energy change expected in similar flyby encounters analysed in their work (for our model we obtain $\beta=$0.8 and $p=$4, see $r^{1/4}$ model in Figs. 2 and 4 in Aguilar \& White (\cite{agwi85})). The corresponding $\Delta M/M$ is about 4\%, which is well within our M/L uncertainty: this means that our assumption of null mass loss is consistent with our model precision. Similar conclusions are found by comparing our results with Funato \& Makito (\cite{fuma99}) : in this case the expected $\Delta M/M$ is about 5\% (for $V/\sigma$=3 and $p/v_\mathrm{r}=$2 for their Hernquist model, Fig. 11 c) and d)).\\
Eq. (\ref{delE}) is the most direct way to estimate the injected energy, but not the most accurate. Aguilar \& White (\cite{agwi85}) pointed out that this ``distant encounter'' approximation can overestimate the real amount of energy change for a large range of impact parameters. From inspection of their Fig. 4, however, we can expect some effect for our encounter geometry : $\beta=$0.8 and $p=$4 imply an overestimate of $\sim 0.02$ in $\Delta E/E$. We will consider this source of uncertainty in the following discussion about the error budget of this analysis.\\
In conclusion, we stress that, within the flyby encounter scenario, we have neglected possible effects of the relative nearby system, NGC~1380. This galaxy has a projected distance of 38$'$, an equally large relative velocity, V$'$=452 km s$^{-1}$, with respect NGC 1399 and is as bright as NGC1404 and possibly as massive. This large projected distance, anyway, would produce a negligible $\Delta E'$ in a present tangential encounter (a flyby encounter in the past cannot be excluded, as suggested by Kissler-Patig et al. \cite{kisl99}).
    
\subsubsection{Recovering $\sigma_\mathrm{obs}$: consistency with the non-equilibrium scenario}
We aim at verifying whether the velocity dispersion profile in the outer 
regions of NGC~1399 can be explained in terms of Eq. (\ref{efin}): we will do it for the last Z1 bin, $Z_\mathrm{last}=182''$.\\
Using the value for $\Delta E$ in the definition of the post-encounter energy, 
$E_\mathrm{fin}$ we obtain 
\begin{equation}
\frac{3}{2}M\sigma_\mathrm{obs}^2=\frac{1}{2}MV_\mathrm{rot}^2+\frac{3}{2}M \sigma_\mathrm{eq}^2+ \Delta E.
\end{equation}
$M$ is the mass content of the region where the injected energy is redistributed which, in our case, is the last bin along the Z1 axis.
Solving with respect to $\sigma_\mathrm{obs}^2$, we have 
\begin{equation}
\sigma_\mathrm{obs}^2=\frac{1}{3}V_\mathrm{rot}^2+\sigma_\mathrm{eq}^2+\frac{2}{3}\frac{\Delta E}{M}.
\label{sobs}
\end{equation}
Using the values of $V_\mathrm{rot}$ and $\sigma_\mathrm{eq}$ extrapolated from the
equilibrium configuration before the encounter, Eq. (\ref{sobs}) gives $\sigma_\mathrm{obs}=264\pm43$ km  s$^{-1}$, 
which is consistent with the estimates from the 
kinematical data for both PNe and GCs (within the errors at $Z_\mathrm{last}$). Here the quoted errors are obtained by propagating the 1$\sigma$ model uncertainties on $\sigma_{eq}$ ($\sim$ 35 km  s$^{-1}$), $M(r)$ ($\delta$M/M=0.3) and $V_{rot}$ ($\sim$ 10 km  s$^{-1}$) at $Z_{\mathrm{last}}$ considered as dependent quantities. This result is consistent, within the estimated error, with the possible overestimate introduced using Eq. (\ref{delE}). In this sense our $\Delta E$ estimate can be considered reasonably accurate.\\   
Larger uncertainties come from our assumption on the encounter geometry. 
The tangential configuration is neither the most obvious, nor the most favorable: it is not difficult to envisage configurations where the relative velocity between the systems is lower than their relative velocity along the line of sight and/or their projected distance is larger than their distance at closest approach. In this latter case the amount of injected energy increases together with$\sigma_\mathrm{obs}$. 
\begin{figure}
\centering
\epsfig{figure=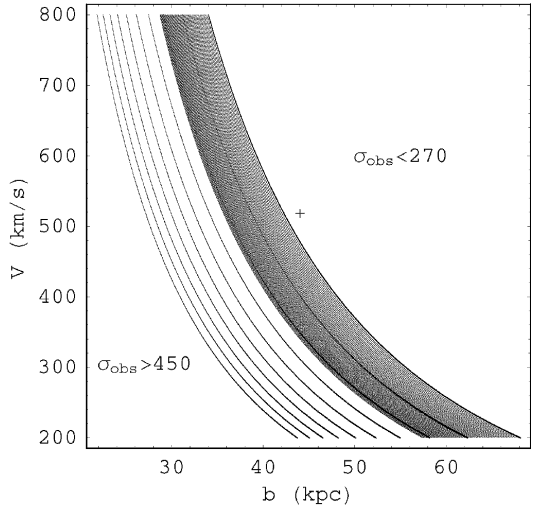,width=7cm,height=7cm}
\caption{\footnotesize $\sigma_\mathrm{obs}$ as a function of $b$ and $V$: contours are from 270 km s$^{-1}$ to 450 km s$^{-1}$ with a step of 20 km s$^{-1}$, the cross is the location of the $\sigma_\mathrm{tan}$ estimate. See the discussion in the text.}
\label{sirange}
\end{figure}
In Fig. \ref{sirange}, $\sigma_\mathrm{obs}$ from Eq. (\ref{sobs}) is shown as a function of $b$ and $V$: the contours cover the $\pm$1RMS range of the $\sigma_\mathrm{last}$ in the last bin, the cross is our tangential estimate, $\sigma_\mathrm{tan}$, and the grey region is the overlap between $\pm$1RMS values of both $\sigma_\mathrm{last}$ and $\sigma_\mathrm{tan}$. The tangential estimate should be considered as an indicative result, because a wide range of configurations can match the observed dispersion. Conclusive results can be obtained by acquiring information on the geometry of the encounter. This uncertainty can be solved with more extended surveys of PNe in the NGC~1399-NGC~1404 system which can possibly disclose the orbit of an ongoing encounter (Moore et al. \cite{moo96}).\\    
Finally, if our assumptions about the inner equilibrium are correct, this model has two consequences: 1) the mass distribution given by the equilibrium before the interaction (Fig. \ref{noeqmo}) is a reasonable description 
of the actual mass content in NGC~1399, since the high-speed encounter 
does not imply a mass modification; 2) the observed kinematics can be explained in term of an energy injection rather than a pure pressure gradient needed to balance a steeply increase in the mass-to-light ratio.
The non-equilibrium hypothesis leads to an outer $M/L$ ratio 
for NGC~1399 which is consistent with the typical values for giant E galaxies obtained using the stellar (including PNe) and GCs kinematics when compared at the same galactocentric distances (NGC5128: Harris et al. \cite{haha88}; Hui et al. \cite{hui95}; NGC1316: Arnaboldi et al. \cite{arn98}, Goudfrooij et al. \cite{goud01}; NGC3115: Emsellem et al. \cite{emsel99}; see also Magorrian \& Ballantyne \cite{maba01} for a collection of 18 early-type galaxies). 

\section{Discussion and Conclusions}
So far the discrete radial velocity fields of PNe and GCs are the only viable diagnostics to extend the dynamical studies into the outer regions of spheroidal galaxies. Furthermore PNe, being associated to the stellar population, allow us to extend the stellar kinematics out to the halo regions. Despite the small number statistics, PNe and GCs can be used efficiently for ``normal'' systems, while they suffer some limitations for the ``disturbed'' systems, where a more detailed kinematical information is needed. In this latter case, we can still hope to derive some indications of the actual dynamical state of their parent galaxies.\\
This is possibly the case of NGC~1399, the cD galaxy in the center of the Fornax cluster. Here, the re-analysis of the PNe data from Arnaboldi et al. (\cite{arn94}), by mean of the procedure tested in Napolitano et al. (\cite{napo01}), shows a peculiar rotational structure: $V_\mathrm{rot}$ has a peak of $\sim250$ km s$^{-1}$ at $R\sim140''$ and then declines. At the same distance from the center, the velocity dispersion has a minimum: it decreases out to $R=140''$ and then increases. 
This could be a kinematical signature of dynamical substructures (a truncated disk, for example) in the inner regions which is not highlighted in published photometrical studies. Anyway, the fact that GCs show the same dispersion profile as the PNe, even if they are a decoupled population (they do not rotate in the regions where they overlap with the PNe data), make this multicomponent scenario inappropriate to explain the observed ``heating'' of the outer  kinematics. This heating possibly calls for a different dynamical justification.\\
In a conventional way, assuming equilibrium, we derived the mass distribution of NGC~1399 accounting for the observed kinematics and found a large dark-mass content with  M/L=56 (Model I) at R=400$''$, fully consistent with previous analyses.
In a different approach, the peculiar kinematics is interpreted as the signature of an interaction situation. Such an interaction scenario has been invoked in previous work to account for the peculiarities of this system: the extended stellar envelope (Schombert \cite{scho86}), the unusual overabundance of globular clusters (Kissler-Patig et al. \cite{kisl99} and references therein), and their metallicity distribution (Minniti et al. \cite{minn98}).
The interaction scenario is also compatible with X-ray observations: Paolillo et al. (\cite{paol01}) have recently analysed new data 
from ROSAT HRI deep observations and found 
features (holes, filamentary, shoulders in the density profile) which are qualitatively expected in a tidal interaction scenario as shown by analytical works (D'Ercole et al. \cite{derc00}) and by N-body simulations (Barnes \cite{bar00}).
These independent evidences of ongoing interaction make any equilibrium analyses in the outer regions of this galaxy quite uncertain: in order to be conservative, this is done, however, in Model II, where we find M/L=33 at R=400$''$.\\
Here we have presented a simplified model as an attempt to approach non-equilibrium situations in a quantitative way. The lack of a detailed kinematics implies assumptions on the geometry of the encounter which affect the final result and therefore the dynamical situation of NGC~1399 requires further investigations. However, this simple dynamical analysis gives two interesting indications: 1) the observed kinematics is coherent with a scenario of non-equilibrium due to a galaxy-galaxy flyby encounter; 2) by considering non-equilibrium, the total mass content required within R=400$''$ is $\sim$30\% less than for the equilibrium model (Model I) and implies M/L=26.
\begin{acknowledgements}
The authors are grateful to the referee, Dr. E. Emsellem, and to O. Gerhard and M. Dopita for their useful comments and suggestions. N.R.N. is receiving financial support from the European Social Found.
\end{acknowledgements}


\begin{thebibliography}{}
\bibitem[1999]{ada99} Adami, C., Marcelin, M., Amram, P., \& Russeil, D., 1999, A\&A 349, 812 
\bibitem[1985]{agwi85} Aguilar, L.A., \& White, S.D.M., 1985, ApJ 295, 374
\bibitem[1994]{arn94} Arnaboldi, M., Freeman, K.C., Capaccioli, M., \& Ford, H., 1994, ESO Messenger 76, 40
\bibitem[1996]{arn96}Arnaboldi, M., Freeman, K.C., Mendez, R., et al. 1996, ApJ 472, 145
\bibitem[1998]{arn98}Arnaboldi, M., Freeman, K.C., Gerhard, O., et al. 1998, ApJ 507, 759
\bibitem[2000]{bar00} Barnes, J.E., 2000, astro-ph/0010145
\bibitem[1994]{bend94}Bender, R., Saglia, R.P., Gerhard, O.E., 1994, MNRAS 269, 785
\bibitem[1989]{bick89} Bicknell, G.V., Carter, D., Killeen, N.E.B., \& Bruce, T.E.G., 1989, ApJ 336, 639B 
\bibitem[1987]{bintr} Binney, J.J., \& Tremaine, S., 1987, Galactic Dynamics, Princeton Series in Astrophysics
\bibitem[1993]{capa93} Capaccioli, M., Cappellaro, E., Held, E.V., \& Vietri, M., 1993, A\&A 274, 69
\bibitem[1989]{ciar89}Ciardullo, R., Jacoby, G.H., Ford, H.C., \& Neill, J.D., 1989, ApJ 339, 53
\bibitem[1991]{ciar91} Ciardullo, R., Jacoby, G.H., \& Harris, W.E., 1991, ApJ 383, 487
\bibitem[1991]{devau91} de Vaucouleurs et al. 1991, Third reference catalogue of bright galaxies
\bibitem[2001]{dale01} Dale, D.A., Giovanelli, R.,
 Haynes, M.P., Hardy, E., \& Campusano, L.E.,  2001, AJ 121, 1886
\bibitem[2000]{derc00} D'Ercole, A., Recchi, S., \& Ciotti, L., 2000, ApJ 533, 799
\bibitem[1995]{dono95} D'Onofrio, M., Zaggia, S.R.,Longo, G., Caon, N., \& Capaccioli, M., 1995, A\&A 296, 319
\bibitem[1980]{dres80} Dressler, A., 1980, ApJ 236, 351
\bibitem[1997]{dres97} Dressler, A. et al., 1997, ApJ 490, 577
\bibitem[1999]{emsel99} Emsellem, E., Dejonghe, H., Bacon, R., 1999, MNRAS, 303, 495
\bibitem[1998]{forb98} Forbes, D.A., Grillmair, C.J., Williger, G.M., Elson, R.A.W., \& Brodie, J.P., 1998, MNRAS 293, 325 
\bibitem[1996]{ford96} Ford, H. C., Hui, X., Ciardullo, R., Jacoby, G. H., \& Freeman, K. C., 1996, ApJ 472, 145
\bibitem[1999]{fuma99} Funato, Y., \& Makito, J., 1999, AJ 511, 625
\bibitem[2001]{goud01} Goudfrooij, P., Mack, J., Kissler-Patig, M., Meylan, G., \& Minniti, D., 2001, MNRAS, 322, 643
\bibitem[1994]{grill94} Grillmair, C.J., Freeman, K.C., Bicknell, G.V., et al. 1994, ApJ 422, 9
\bibitem[1998]{grah98} Graham, A. W., Colless, M. M., Busarello, G., Zaggia, S., \& Longo, G. 1998, A\&AS 133, 325
\bibitem[1988]{haha88} Harris, H.C., Harris, G.L.H., \& Hesser, J.E., 1988, IAU Symp. 126, 545
\bibitem[1984]{hay84} Haynes, M.P., Giovanelli, R., \& Chincarini, G.L., 1984, ARA\&A 22, 445
\bibitem[1995]{hui95} Hui, X., Ford, H., Freeman, K.C., \& Dopita, M.A., 1995, ApJ 449, 592
\bibitem[1996]{ike96} Ikebe, Y., et al., 1996, Nature 379, 427
\bibitem[1996]{irw96} Irwin, J.A., \& Sarazin, C.L., 1996, ApJ 471, 683
\bibitem[1997]{jon97} Jones, C., Stern, C., Forman, W., et al. 1997, ApJ 482, 143
\bibitem[1999]{kauf99} Kauffmann, G., 1999, AAS 195, 6701 
\bibitem[1988]{kill88} Killeen, N.E.B., \& Bicknell, G.V.,  1988, ApJ 325, 165
\bibitem[1999]{kisl99} Kissler-Patig, M., Grillmair, C.J., Meylan , G., et al. 1999, AJ 117, 1206
\bibitem[2000]{kiku00} Kikuchi, K., et al., 2000, ApJ 531, 95
\bibitem[2001]{maba01} Magorrian, J., \& Ballantyne, D., 2001, MNRAS 322, 702
\bibitem[1993]{mcmil93} McMillan, R., Ciardullo, R. \& Jacoby, G.H., 1993, ApJ 416, 62
\bibitem[1997]{merr97} Merritt, D., 1997, AJ 114, 228
\bibitem[2000]{miho00} Mihos, J. C., 2000, ASP Conference Series 197, 275
\bibitem[1998]{minn98} Minniti, D., Kissler-Patig, M., Goudfrooij,P., \& Meylan, G., 1998, AJ 115, 121
\bibitem[1996]{moo96} Moore, B., Katz, N., Lake, G., Dressler, A., \& Oemler, A.Jr., 1996, Nature 379, 613
\bibitem[2001]{napo01} Napolitano, N.R., Arnaboldi, M., Freeman, K.C., \& Capaccioli, M., 2001, A\&A, 377, 784 (NAFC)
\bibitem[1991]{niet91} Nieto, J.L., Bender, R., Arnaud, J., \& Surma, P., 1991, A\&A 244, 25
\bibitem[2001]{paol01} Paolillo, M., Fabbiano, G., Peres, G., \& Kim, D.W., 2001, ApJ, accepted, astro-ph/0109342
\bibitem[1995]{rang95} Rangarajan, F.V.N., Fabian, A.C., Forman, W.R., \& Jones, C., 1995, MNRAS 272, 665
\bibitem[1995]{roet95} Roettiger, K., Burns, J.O., \& Pinkney, J., 1995, ApJ 453, 634
\bibitem[1996]{roet96} Roettiger, K., Burns, J.O., \& Loken, C., 1996, ApJ 473, 651
\bibitem[1999]{rub99} Rubin, V.C., Waterman, A.H., \&
 Kenney, J.D.P., 1999, AJ 118, 236
\bibitem[2000]{sagl00} Saglia, R. P., Kronawitter, A., Gerhard, O., \& Bender, R., 2000, AJ 119, 153
\bibitem[2000]{salau00} Salo, H. \& Laurikainen, E., 2000, MNRAS 319, 377 
\bibitem[1986]{scho86} Schombert, J.M., 1986, ApJS 60, 603  
\bibitem[1998]{scor98} Scorza, C., Bender, R., Winkelmann, C., Capaccioli, M., \& Macchetto, D.F., A\&AS 131, 265
\bibitem[2001]{sola01} Solanes, J.M., Manrique, A., García-Gómez, C., et al. 2001, ApJ 548, 97
\bibitem[2000]{suka00} Sugerman, B., Summers, F.J., \& Kamionkowski, M., 2000, MNRAS 311, 762 
\bibitem[1994]{vdbor94} van den Bosch, F.C., Ferrarese, L., Jaffe, W., Ford, H.C., \& O'Connell, R.W., 1994, AJ 108, 1579
\bibitem[1998]{vdbor98} van den Bosch, F.C., 1998, ApJ 507, 601
\bibitem[1988]{whit88} Whitmore, B.C., Forbes, D.A., \& Rubin, V.C., 1988, ApJ 333, 542
\bibitem[1993]{whit93} Whitmore, B.C., Gilmore, D.M., \& Jones, C., 1993, ApJ 407, 489 
\bibitem[1993]{wifr93}Winsall, M.L., \& Freeman, K.C., 1993, A\&A 268, 443
\end{thebibliography}
\end{document}